\documentclass[12pt]{article}

\usepackage{amsmath}
\usepackage{pdflscape}
\usepackage{color}
\usepackage{amssymb}
\usepackage{graphicx}
\usepackage{subfigure}
\usepackage{hyperref}
\usepackage{mathrsfs}
\usepackage{latexsym}
\usepackage{amsfonts}
\usepackage{amsmath}
\usepackage{esint}
\usepackage{amsxtra}
\usepackage{longtable}
\usepackage{rotating}
\usepackage{multirow}
\usepackage[titletoc,title]{appendix}

\newdimen\mathindent
\mathindent\leftmargini

\def \ep1{\epsilon_1}
\def \ep2{\epsilon_2}
\def \m{\mbox}

\def \be{\begin{equation}}
\def \ee{\end{equation}}
\def\beq{\begin{eqnarray}}
\def\eeq{\end{eqnarray}}
\def \ba{\begin{array}}
\def \ea{\end{array}}

\def \f{\frac}

\def \p{\partial}

\def \sn{\mbox{sn}}
\def \cn{\mbox{cn}}
\def \dn{\mbox{dn}}
\def \ep{\epsilon}

\begin{document}
\vspace*{-.6in}
\thispagestyle{empty}
\baselineskip = 18pt

\vspace{.5in}
\vspace{.5in}
{\LARGE
\begin{center}
A new treatment for some periodic Schr\"{o}dinger operators II: the wave function
\end{center}}

\vspace{1.0cm}

\begin{center}
\renewcommand{\thefootnote}{\fnsymbol{footnote}}
Wei He\footnote[1]{weihephys@foxmail.com}
\setcounter{footnote}{0}
\vspace{1.0cm}\\
\em{School of Electronic Engineering,\\Chengdu Technological University, Chengdu 611730, China}

\end{center}
\vspace{1.0cm}

\begin{center}
\textbf{Abstract}
\end{center}
\begin{quotation}
\noindent  Following the approach of our previous paper we continue to study the asymptotic solution of periodic Schr\"{o}dinger operators. Using the eigenvalues obtained earlier the corresponding asymptotic wave functions are derived. This gives further evidence in favor of the monodromy relations for the Floquet exponent proposed in the previous paper. In particular, the large energy asymptotic wave functions are related to the instanton partition function of N=2 supersymmetric gauge theory with surface operator. A relevant number theoretic dessert is appended.
\end{quotation}




\pagenumbering{arabic}

\newpage

\tableofcontents

\section{Introduction}

In the previous paper Ref.~\cite{wh1412} we study how the Floquet theory manifests in the multiple asymptotic spectral solutions of some periodic Schr\"{o}dinger operators. We have only studied the eigenvalue aspect of these solutions. Following this method, it is very convenient to derive the corresponding asymptotic wave functions, we present the results in this paper. We focus on our canonical examples: the Mathieu equation and the Lam\'{e} equation, which are the most widely used in periodic spectral problem. Our main conclusion are made for elliptic potentials, but as we have explained, to collect evidences for the proposals a crucial consistent requirement is that the solutions of the Lam\'{e} equation must reduce to corresponding solutions of the Mathieu equation. Therefore the Mathieu equation, which is much better understood, is included here as a reference example.
Unlike in Ref.~\cite{wh1412}, in this paper we do not use the ellipsoidal wave equation as the example for the elliptic potential as it would lead to lengthly formulae for wave functions, instead the wave functions of Lam\'{e} equation are already enough for our purpose.

In the Section \ref{sectionlargeenergywavefunction} we derive the wave functions for large energy (weak coupling) perturbation. In the Section \ref{sectionsmallenergywavefunction} we derive the  wave functions for small energy (strong coupling) perturbation.
Some wave functions has been studied before, we briefly comment the old materials where earlier treatments can be found. In the Section \ref{connectoN=2} we explain how the eigenfunctions are related to supersymmetric gauge theory, in the context of Gaueg/Bethe correspondence \cite{NS0908}.

The conclusion of this paper is that the wave functions give further evidence for the relations between multiple asymptotic solutions and the Floquet property associated with multiple periods. Among the wave functions, the eigenfunction (\ref{lame2ndsmallwf}) is a new solution.

\section{Large energy wave function}\label{sectionlargeenergywavefunction}

\subsection{The Floquet wave function}

As we have shown previously, the large energy perturbation can be carried out using a method from KdV theory \cite{MGK, classintegrable}, the wave function is given by
\begin{equation}
\psi(x)=\exp(\int^xv(y)dy),
\end{equation}
where $v(x)$ satisfies $v_x+v^2=u+\lambda$ and has the WKB expansion
\begin{equation}
v(x)=\sqrt{\lambda}+\sum_{\ell=1}^{\infty}\f{v_\ell(x)}{(\sqrt{\lambda})^\ell}.\label{nuexpansion}
\end{equation}
$v_\ell(x)$ are given by the KdV Hamiltonian densities \cite{classintegrable}.

In fact there is another sector of solution for the $v(x)$, it is expanded in the same form of (\ref{nuexpansion}) with $\sqrt{\lambda}$ changed to $-\sqrt{\lambda}$. Therefore we get the unnormalized asymptotic expansion of the corresponding wave functions,
\begin{equation}
\psi_\pm(x)=\exp\Big(\pm\lambda^{\f{1}{2}}x
\pm\f{1}{2\lambda^{\f{1}{2}}}\int^xu(y)dy
-\f{1}{4\lambda}u(x)\mp\f{1}{8\lambda^{\f{3}{2}}}[\int^xu^2(y)dy-u_x]
+\cdots\Big).\label{largeenergywavefunction}
\end{equation}
The dispersion relation $\lambda(\nu)$, where $\nu$ is the Floquet exponent, is obtained by the classical Floquet theory,
\begin{equation}
\exp\Big(\int_x^{x+T}v(y)dy\Big)=\exp(\pm i\nu T).\label{largeenegyfloquet}
\end{equation}
The relations (\ref{largeenergywavefunction}) and (\ref{largeenegyfloquet}) give a complete perturbative solution for large energy.

\subsection{Mathieu equation}

The potential for the Mathieu equation is
\begin{equation}
u(x)=2h\cos2x,\label{mathieupotential1}
\end{equation}
from (\ref{largeenergywavefunction}) we get the large energy asymptotic wave functions
\begin{equation}
\psi_\pm(x)=\exp\Big(\pm\lambda^{\f{1}{2}}x\pm\f{h\sin2x}{2\lambda^{\f{1}{2}}}
-\f{h\cos2x}{2\lambda}
\mp\f{8h\sin2x+4h^2x+h^2\sin4x}{16\lambda^{\f{3}{2}}}+\cdots\Big).\label{psilambdamathieu}
\end{equation}
It is clear that with $\lambda$ as the expansion parameter, the coefficients of $\lambda^{-\f{\ell}{2}}$ are not necessary periodic functions. However, the wave functions must satisfy the Floquet property, this is made clear by the following parameter change using the eigenvalue expansion.

The large energy dispersion relation is a classic result, see for example Refs.~\cite{mclachlan, Arscott1964, Wang-Guo, NIST}. The relation (\ref{largeenegyfloquet}) can be used to compute, it is
\begin{equation} \lambda=-\nu^2-\f{h^2}{2\nu^2}-\f{h^2}{2\nu^4}-\f{16h^2+5h^4}{32\nu^6}+\cdots.\label{largeenergyeigenvaluemathieu}
\end{equation}
It has a solution
\begin{equation}
\lambda^{\f{1}{2}}=i(\nu+\f{h^2}{4\nu^3}+\f{h^2}{4\nu^5}+\f{16h^2+3h^4}{64\nu^7}+\cdots).
\end{equation}
Another solution $-\lambda^{\f{1}{2}}$ does not lead to new wave function. Substitute $\lambda^{\f{1}{2}}$ into (\ref{psilambdamathieu}) we get the wave functions in the form
\begin{equation}
\psi_\pm(x)=\exp\Big(\pm i\nu x\mp\f{ih\sin2x}{2\nu}+\f{h\cos2x}{2\nu^2}\mp\f{i(8h\sin2x+h^2\sin4x)}{16\nu^3}+\cdots\Big).
\label{largeenergywavefunctionmathieu}
\end{equation}
Now the coefficients of $\nu^{-l}$, with $l\geqslant1$, are periodic functions, the wave functions take the form $\psi_\pm(x)=e^{\pm i\nu x}\phi(\pm x)$ with $\phi(x)$ a periodic function.
This wave function is related to the N=2 pure Yang-Mills gauge theory with surface operator, see the discussion in the Section \ref{connectoN=2}.

Another bases of the asymptotic wave functions, commonly used in many literatures, are $ce_m(x)$ and $se_m(x)$. Up to a constant, their relation to $\psi_\pm(x)$ is
\begin{equation}
\psi_\pm(x)|_{\nu=m}\sim ce_m(x)\pm i se_m(x).
\end{equation}

\subsection{Lam\'{e} equation}\label{lamelargeenergy}

The potential for the Lam\'{e} equation is an elliptic function, for the large energy perturbation we should use the Weierstrass form to obtain compact formulae. In this paper we use the potential
\begin{equation}
u(x)=\alpha\widetilde{\wp}(x;2\omega_1,2\omega_2),
\end{equation}
which is defined by a shifted elliptic function $\widetilde{\wp}(x;2\omega_1,2\omega_2)=\wp(x;2\omega_1,2\omega_2)+\zeta_1$ with $\zeta_1$ a constant given  by the Weierstrass zeta function $\zeta_1=\f{\zeta(\omega_1)}{\omega_1}$. The coupling constant $\alpha$ is often represented as $n(n-1)$, the nome of the elliptic function is $q=\exp(2\pi i\f{\omega_2}{\omega_1})$. The use of shifted potential is more convenient for us to see the relation to N=2 supersymmetric gauge theory in the Section \ref{connectoN=2}, it does not change essential aspects of the spectral solution. As we have demonstrated in Ref.~\cite{wh1412} the large energy asymptotic solution is related to the period $2\omega_1$. Then the corresponding wave functions are
\begin{equation}
\psi_\pm(x)=\exp\Big(\pm\lambda^{\f{1}{2}}x\mp\f{\alpha\widetilde{\zeta}(x)}{2\lambda^{\f{1}{2}}}-\f{\alpha\widetilde{\wp}(x)}{4\lambda}   \mp\f{2\alpha(\alpha-6)\widetilde{\wp}_x-24\alpha^2\zeta_1\widetilde{\zeta}(x)+\alpha^2(g_2-12\zeta_1^2)x}{96\lambda^{\f{3}{2}}}+\cdots\Big),\label{psilambdalame}
\end{equation}
where  $g_2, g_3$ are the modular invariants of $\wp(x)$. We use another shifted function defined by $\widetilde{\zeta}(x)=\zeta(x)-\zeta_1x$, it is $2\omega_1$-periodic and satisfies the relation $\p_x\widetilde{\zeta}(x)=-\widetilde{\wp}(x)$. Again, some coefficients of $\lambda^{-\f{n}{2}}$ are not periodic functions.

The corresponding dispersion relation was derived by E. Langmann \cite{Langmann2004b}, expressed as a $q$-series, the same expression also appears in the context of its relation to gauge theory \cite{NS0908}, we examined this relation in Ref.~\cite{wh1108}. Another way to derive the dispersion relation is to use the formula (\ref{largeenegyfloquet}), then we get an expression involving quasi-modular functions \cite{wh1401},
\begin{equation}
\lambda=-\nu^2+\f{\alpha^2(12\zeta_1^2-g_2)}{48\nu^2}+\f{\alpha^3(20\zeta_1^3-g_2\zeta_1-g_3)-\alpha^2(2g_2\zeta_1-3g_3)}{80\nu^4}+\cdots.\label{largeenergyeigenvaluelame}
\end{equation}
It has a solution
\begin{equation}
\lambda^{\f{1}{2}}=i\Big(\nu-\f{\alpha^2(12\zeta_1^2-g_2)}{96\nu^3}-\f{\alpha^3(20\zeta_1^3-g_2\zeta_1-g_3)-\alpha^2(2g_2\zeta_1-3g_3)}{160\nu^5}+\cdots\Big).
\end{equation}
Substitute $\lambda^{\f{1}{2}}$ into (\ref{psilambdalame}) we get the wave functions in the Floquet form,
\begin{equation}
\psi_\pm(x)=\exp\Big(\pm i\nu x\pm\f{i\alpha\widetilde{\zeta}(x)}{2\nu}+\f{\alpha\widetilde{\wp}(x)}{4\nu^2}\pm\f{i[12\alpha^2\zeta_1\widetilde{\zeta}(x)-\alpha(\alpha-6)\widetilde{\wp}_x]}{48\nu^3}+\cdots\Big).
\label{largeenergywavefunctionlame}
\end{equation}
The wave functions also satisfy the property $\psi_\pm(-x)=\psi_\mp(x)$. In the Section \ref{connectoN=2} we would show the connection of this wave function and the partition function of the N=$2^*$ supersymmetric gauge theory with surface operator.

There is a comment about the polynomials of elliptic functions that appear in (\ref{psilambdalame}) and (\ref{largeenergywavefunctionlame}). Recall that any elliptic function can be expressed as a linear combination of zeta functions and their derivatives. In fact, the Hamiltonian densities $v_\ell(x)$ for the elliptic potential have no pole of order one at $x=0$, therefore, they are linear combinations of  $\p_x^k\widetilde{\wp}(x)$ with $k\geqslant0$. Then the integrated Hamiltonians appearing in (\ref{psilambdalame}) are linear combinations of $x$ and $\p_x^k\widetilde{\zeta}(x)$ with $k\geqslant0$.
In the wave function (\ref{largeenergywavefunctionlame}) the phase $e^{\pm i\nu x}$ contains the linear term of $x$, the coefficients of $\nu^{-l}$ are linear combinations of $\p_x^k\widetilde{\zeta}(x)$ with $k\geqslant0$, probably include a $x$-independent constant term. The constant terms can be absorbed into the normalization constant, then the expressions are linear polynomials of $\p_x^k\widetilde{\zeta}(x)$ with $k\geqslant0$. Or equivalently, because $\widetilde{\zeta}(x)=\p_x\ln\vartheta_1(\f{\pi x}{2\omega_1},q)$, they are linear polynomials of $\p_x^k\ln\vartheta_1(\f{\pi x}{2\omega_1},q)$ with $k\geqslant1$. This point is important when we connect the wave function to the instanton partition function in the Section \ref{connectoN=2}, especially for higher order terms which we do not explicitly give in (\ref{largeenergywavefunctionlame}).

When take the limit $q\to0, \alpha\to\infty$ with $\alpha q^{\f{1}{2}}\to-\f{h}{4}$ fixed, all the results obtained here reduce to the case of the Mathieu equation. Some details in the limit process need further explanation. We first examine how the elliptic potential is reduced to the trigonometric potential. Instead of taking the limit for $\widetilde{\wp}(x)$, we have to shift the argument and take the limit for $\widetilde{\wp}(x+\omega_2)$. From the leading order expansion of $\lim\limits_{q\to0}\widetilde{\wp}(x+\omega_2)$ given below in (\ref{limitofwp}), the resultant potential takes the form
\begin{equation}
u(x)=2h(\f{\pi}{2\omega_1})^2\cos\f{\pi x}{\omega_1},\label{mathieupotential2}
\end{equation}
with period $2\omega_1$, its eigenvalue denoted by $\widetilde{\lambda}$ can be obtained from the limit of (\ref{largeenergyeigenvaluelame}). If we use the rescaled coordinate $\chi=\f{\pi x}{2\omega_1}$ and eigenvalue $\lambda=(\f{2\omega_1}{\pi})^2\widetilde{\lambda}$, the limit of the Lam\'{e} equation could be written in the standard form of the Mathieu equation,
\begin{equation}
(\p_\chi^2-2h\cos2\chi)\psi=\lambda\psi.
\end{equation}
The corresponding large energy asymptotic eigenvalue $\lambda$ and wave functions $\psi_\pm(\chi,\nu)$ take the same functional form as the eigenvalue (\ref{largeenergyeigenvaluemathieu}) and  the wave functions (\ref{largeenergywavefunctionmathieu}), but with the coordinate variable substituted by $\chi$.

Let us inspect more carefully the limit for eigenvalue and wave functions of the Lam\'{e}  equation.
For the wave functions (\ref{largeenergywavefunctionlame}), we shift the argument by $x\to x+\omega_2$, then take the limit $q\to0$(i.e. with $\omega_1$ fixed, $\omega_2\to i\infty$) for $\psi_\pm(x+\omega_2)$. The following expansions are needed,
\begin{align}
&\lim_{q\to0}\widetilde{\wp}(x+\omega_2)=(\f{\pi}{2\omega_1})^2\Big(-8q^{\f{1}{2}}\cos2\chi-16q\cos4\chi-8q^{\f{3}{2}}(\cos2\chi+3\cos6\chi)+\mathcal {O}(q^2)\Big),\label{limitofwp}\\
&\lim_{q\to0}\widetilde{\zeta}(x+\omega_2)=\f{\pi}{2\omega_1}\Big(4q^{\f{1}{2}}\sin2\chi+4q\sin4\chi+4q^{\f{3}{2}}(\sin2\chi+\sin6\chi)+\mathcal {O}(q^2)\Big).
\label{limitofzeta}
\end{align}
The Lam\'{e} wave functions (\ref{largeenergywavefunctionlame}) indeed reduce to the Mathieu wave functions, iff we further substitute the Floquet exponent by $\widetilde{\nu}=\f{\pi}{2\omega_1}\nu$.
The modification of the exponent can be understood as follows. In this subsection, the exponent $\nu$ conjugates to the variable $\chi$ of the $\pi$-periodic potential $\cos2\chi$, it is different from the exponent that  conjugates to the variable $x$ of the $2\omega_1$-periodic potential $\cos\f{\pi x}{2\omega_1}$ which we should have denoted by another letter $\widetilde{\nu}$. According to the Floquet theorem their wave functions produce phases under periodic shift by $\psi_\pm(\chi+\pi,\nu)=\exp(\pm i\nu\pi)\psi_\pm(\chi,\nu)$ and $\psi_\pm(x+2\omega_1,\widetilde{\nu})=\exp(\pm i2\widetilde{\nu}\omega_1)\psi_\pm(x,\widetilde{\nu})$. But actually the potentials are the same, so the phases must be the same, which means the exponents are related by the relation $\nu\pi=2\widetilde{\nu}\omega_1$. It is easy to check that the limit for the eigenvalue (\ref{largeenergyeigenvaluelame}) which is associated to the potential $\cos\f{\pi x}{2\omega_1}$, hence with $\nu$ substituted by $\widetilde{\nu}$, equals $(\f{\pi}{2\omega_1})^2\lambda$ where $\lambda$ is the Mathieu eigenvalue for the potential $\cos2\chi$.

\section{Small energy wave function}\label{sectionsmallenergywavefunction}

\subsection{Location of small energy perturbation}

Besides the large energy solution, there exists other solutions which are small energy excitations around local minima, i.e. the critical points of potential.
We notice for some periodic potentials at each local minimum there is an asymptotic solution, and all known asymptotic solutions are located at a local minimum \cite{wh1412, wh1108}.

For example, the potential $u(x)=2h\cos2x$ has local minima at $x_*=0$ and $x_*=\f{\pi}{2}$ modulo periods. At the minima $u(x_*)=\pm2h$, therefore the eigenvalues take the form
\begin{equation}
\lambda=\mp2h+\delta,
\end{equation}
where $\delta$ is the energy of small excitations. The small energy perturbation is also the strong coupling solution for the potential, $h\gg1$, see Refs.~\cite{Arscott1964} (Chapter V). In a similar way, the elliptic potential $u(x)=\alpha\widetilde{\wp}(x;2\omega_1,2\omega_2)$ has local minima at $x_*=\omega_i$, where the potential $u(x_*)=\alpha(e_i+\zeta_1)$, $i=1,2,3$. The first minima at $x_*=\omega_1$ is associated to the large energy excitations already analyzed in the subsection \ref{lamelargeenergy}, the leading order energy comes from the quasimomentum $\lambda\sim-\nu^2+\cdots$. The other two minima are associated to small energy perturbative solutions, nevertheless, in order to get compact formulae we should use the Jacobian form of the Lam\'e equation to compute.

In this section we derive the corresponding strong coupling wave function, they have the Floquet form, and for elliptic potential their monodromies along periods $2\mathbf{K}$ and $2\mathbf{K}+2\mathbf{K}^{'}$ indeed satisfy the relations we proposed in the previous paper \cite{wh1412}.

\subsection{Mathieu equation}

{\bf The first small energy perturbation}

Around the minimum $x_*=0, \lambda=-2h+\delta$, the potential strength $h\gg1$ is large compared to the energy $\delta$, therefore the expansion parameter is $h^{\f{1}{2}}$. The relation $v_x+v^2=u+\lambda$ has an asymptotic solution in the form \cite{wh1412}

\begin{equation}
v(x)=\sum_{\ell=-1}^{\infty}\f{v_\ell(x)}{(\sqrt{h})^\ell},\label{mathieusmallvexp}
\end{equation}
with
\begin{subequations}
\begin{align}
v_{-1}&=2i\sin x, \\
v_0&=-\f{1}{2}\cot x,\\
v_1&=\f{i}{16}\csc x(\cot^2x+2\csc^2x-4\delta),\\
v_2&=\f{1}{32}\csc^2x\cot x(\cot^2x+5\csc^2x-4\delta),\quad \m{etc}.
\end{align}
\end{subequations}
Another solution can be obtained by changing the signs of odd terms $v_{2\ell+1}$ to $-v_{2\ell+1}$; or with $v_\ell$ unaltered but changing $\sqrt{h}$ to $-\sqrt{h}$ in (\ref{mathieusmallvexp}). This leads to the following asymptotic wave functions,
\begin{align}
\psi_\pm(x)=&\exp\Big(\pm2ih^{\f{1}{2}}\cos x-\f{1}{2}\ln\sin x\pm\f{i}{2^5h^{\f{1}{2}}}[(8\delta-1)\ln\tan\f{x}{2}+3\csc x\cot x]\nonumber\\
&-\f{1}{2^7h}(\cot^4x+5\csc^4x-8\delta\csc^2x)+\cdots\Big).
\end{align}

We can change the parameter $\delta$ to the Floquet exponent $\nu$, by the following strong coupling expansion of the dispersion relation which is well known \cite{mclachlan, Arscott1964, Wang-Guo, NIST},
\begin{equation}
\lambda=-2h+4\nu h^{\f{1}{2}}-\f{4\nu^2-1}{2^3}-\f{4\nu^3-3\nu}{2^6h^{\f{1}{2}}}-\f{80\nu^4-136\nu^2+9}{2^{12}h}+\cdots.
\end{equation}
Then we get the wave functions in the form
\begin{align}
\psi_\pm(x)=&(\sin\f{x}{2})^{\pm i\nu-\f{1}{2}}(\cos\f{x}{2})^{\mp i\nu-\f{1}{2}}\exp\Big(\pm i2h^{\f{1}{2}}\cos x+\f{\csc^2x}{2^5h^{\f{1}{2}}}[4\nu\pm i(3-4\nu^2)\cos x+4\nu\cos2x]\nonumber\\
&-\f{\csc^4x}{2^{10}h}[42-124\nu^2\mp i(155\nu-36\nu^3)\cos x+4\cos2x\pm i(3\nu-4\nu^3)\cos3x\nonumber\\&
+2(1-2\nu^2)\cos4x]+\cdots\Big).
\label{mathieu1stsmallwf}\end{align}
Because the wave functions are unnormalized, the terms in the exponent might appear in slightly different form, nevertheless the differences are constants and can be absorbed into the normalization constant. This comment applies to all of the asymptotic wave functions in this paper.
When the exponent $\nu$ takes real value the asymptotic wave functions have the property $\psi_\pm^*(x)=\psi_\mp(x)$. This solution seems often not recorded in the mathematical literature, however, it was analyzed in a paper by M. Stoner and J. Reeve \cite{Stone-Reeve1978}. The book Ref.~\cite{MullerKirstenQUANT} contains a discussion about this solution in the context of quantum mechanics.

\vspace{3mm}
{\bf The second small energy perturbation}

Around the minimum $x_*=\f{\pi}{2}, \lambda=2h+\delta$, the potential strength $h^{\f{1}{2}}$ again serves as the expansion parameter. The relation $v_x+v^2=u+\lambda$ has an asymptotic solution in the form (\ref{mathieusmallvexp}), with
\begin{subequations}
\begin{align}
v_{-1}&=2\cos x, \\
v_0&=\f{1}{2}\tan x,\\
v_1&=-\f{1}{16}\sec x(\tan^2x+2\sec^2x-4\delta),\\
v_2&=\f{1}{32}\sec^2x\tan x(\tan^2x+5\sec^2x-4\delta),\quad \m{etc}.
\end{align}
\end{subequations}
As in the previous solution, the other solution is obtained by changing $\sqrt{h}$ to $-\sqrt{h}$. The corresponding wave functions are
\begin{align}
\psi_\pm(x)=&\exp\Big(\pm2h^{\f{1}{2}}\sin x-\f{1}{2}\ln\cos x\pm\f{1}{2^5h^{\f{1}{2}}}[(8\delta-1)\ln\f{\cos\f{x}{2}+\sin\f{x}{2}}{\cos\f{x}{2}-\sin\f{x}{2}}-3\sec x\tan x]\nonumber\\
&+\f{1}{2^7h}(\tan^4x+5\sec^4x-8\delta\sec^2x)+\cdots\Big).
\end{align}

The dispersion relation at this local minimum \cite{mclachlan, Arscott1964, Wang-Guo, NIST}
\begin{equation}
\lambda=2h-4\nu h^{\f{1}{2}}+\f{4\nu^2+1}{2^3}+\f{4\nu^3+3\nu}{2^6h^{\f{1}{2}}}+\f{80\nu^4+136\nu^2+9}{2^{12}h}+\cdots
\end{equation}
allows us to change the parameter $\delta$ to the Floquet exponent $\nu$. Then we get the corresponding  asymptotic wave functions
\begin{align}
\psi_\pm(x)=&(\sec x)^{\nu+\f{1}{2}}\Big[
\begin{tabular}{l}
$\cos$\cr
$\sin$
\end{tabular}
(\f{x}{2}+\f{\pi}{4})\Big]^{2\nu}
\exp\Big(\pm 2h^{\f{1}{2}}\sin x+\f{\sec^2x}{2^5h^{\f{1}{2}}}[8\nu\mp(3+4\nu^2)\sin x]\nonumber\\
&+\f{\sec^4x}{2^{10}h}[39+112\nu^2\mp(155\nu+36\nu^3)\sin x\mp(3\nu+4\nu^3)\sin3x\nonumber\\
&-8(1+2\nu^2)\cos2x+\cos4x]+\cdots\Big).
\label{mathieu2ndsmallwf}\end{align}
The study of this solution dates back to the work of E. Ince and the the work of S. Goldstein in the 1920s. Some recent materials easier to access include the paper by R. Dingle and H. M\"{u}ller \cite{Dingle-Muller1962}, the books by N. W. McLachlan \cite{mclachlan}, by F. Arscott \cite{Arscott1964} and by H. M\"{u}ller-Kirsten \cite{MullerKirstenQUANT}.

This asymptotic solution is related to the large $h$ limit of the standard Mathieu functions by
\begin{equation}
(\psi_++\psi_-)|_{\nu=m+\f{1}{2}}\sim ce_m(x),\qquad (\psi_+-\psi_-)|_{\nu=m+\f{1}{2}}\sim se_{m+1}(x),
\end{equation} with $m$ takes either even or odd integers \cite{Arscott1964}. As we have $\psi_\pm(-x)=\psi_\mp(x)$,
then $ce_m(x)$ is an even function and $se_m(x)$ is an odd function, as desired.

\subsection{Lam\'{e} equation}

Now we turn to the more interesting case of elliptic potential where the advantage of our method becomes more transparent. As we have shown in Refs.~\cite{wh1108, wh1412}, for small energy perturbative solution the Jacobian form of the elliptic function is more suitable, therefore we rewrite the potential as $u(z)=\alpha k^2\sn^2(z|k^2)$, and the Lam\'{e} equation is
\begin{equation}
(\p_z^2-\alpha k^2\sn^2z)\psi=\Lambda\psi.
\end{equation}
To transform the elliptic functions from $\widetilde{\wp}(x)$ to $\sn^2z$ we use the following relations of the coordinates, eigenvalues, the elliptic modulus and the nome,
\begin{equation}
x=\f{z+i\mathbf{K}^{\prime}}{(e_1-e_2)^{1/2}},\qquad \lambda=(e_1-e_2)\Lambda-(e_2+\zeta_1)\alpha,\qquad k^2=\f{\vartheta_2^4(q)}{\vartheta_3^4(q)}.
\end{equation}
We also use $\mu$ to denote the Floquet exponent throughout of this subsection, it is different from the Floquet exponent $\nu$ used for the Weierstrass form \cite{wh1108}.

The locations of the small energy perturbations are given by two solutions of the condition  $\p_z\sn^2z=0$ at $z_*=0$ and $z_*=\mathbf{K}$ which correspond to $u(z_*)=0$ and $u(z_*)=\alpha k^2$.

\vspace{3mm}
{\bf The first small energy expansion}

Around $z_*=0$ we have $\Lambda=0+\m{\em{small correction}}$, therefore in this case $\Lambda$ itself is a small quantity compared with the potential strength $\alpha k^2$. Therefore $\frac{1}{\sqrt{\alpha}}$ serves as the expansion parameter. The relation $v_z+v^2=u+\Lambda$ has an asymptotic solution that can be expanded as
\begin{equation}
v(z)=\sum_{\ell=-1}^{\infty}\f{v_\ell(z)}{(\sqrt{\alpha})^\ell},
\end{equation}
with
\begin{subequations}
\begin{align}
v_{-1}&=k\,\sn\,z, \\
v_0&=-\f{1}{2}\p_z\ln\sn\,z,\\
v_1&=\f{1}{8}k\,\sn\,z+\f{4\Lambda+1+k^2}{8k\,\sn\,z}-\f{3}{8k\,\sn^3z},\\
v_2&=-\f{1}{16k^2}\p_z\Big(\f{4\Lambda+1+k^2}{\sn^2z}-\f{3}{\sn^4z}\Big),\quad\m{etc}.
\end{align}
\end{subequations}
The minus sector is obtained by changing $\sqrt{\alpha}$ to $-\sqrt{\alpha}$. Then by a straightforward integration we get the following asymptotic wave functions,
\begin{align}
\psi_\pm(z)=&\exp\Big(\pm\alpha^{\f{1}{2}}\ln(\dn\,z-k\,\cn\,z)-\f{1}{2}\ln\sn\,z\nonumber\\
&\pm\f{1}{2^4\alpha^{\f{1}{2}}k}[\f{3\cn\,z\,\dn\,z}{\sn^2z}+2k\ln(\dn\,z-k\,\cn\,z)-(8\Lambda-1-k^2)\ln\f{\dn\,z+\cn\,z}{\sn\,z}]\nonumber\\
&+\f{3-(4\Lambda+1+k^2)\sn^2z}{2^4\alpha k^2\sn^4z}+\cdots
\Big).
\end{align}

In order to change the parameter $\Lambda$ to the Floquet exponent $\mu$, we use the widely known strong coupling expansion of the dispersion relation \cite{Arscott1964, NIST, MullerKirstenQUANT}
\begin{align}
\Lambda=&-i2\alpha^{\f{1}{2}}k\mu-\f{1}{2^3}(1+k^2)(4\mu^2-1)\nonumber\\
&-\f{i}{2^5\alpha^{\f{1}{2}}k}[(1+k^2)^2(4\mu^3-3\mu)-4k^2(4\mu^3-5\mu)]\nonumber\\
&+\f{1}{2^{10}\alpha k^2}(1+k^2)(1-k^2)^2(80\mu^4-136\mu^2+9)+\cdots.
\label{ellipsoidalJaceigenvalue2}
\end{align}
It leads to the wave functions in the form
\begin{align}
\psi_\pm(z)=&\exp\Big(\pm\alpha^{\f{1}{2}}\ln(\dn\,z-k\,\cn\,z)
-\f{1}{2}(\ln\sn\,z\mp2i\mu\ln\f{\dn\,z+\cn\,z}{\sn\,z})\nonumber\\
&\pm\f{1}{2^4\alpha^{\f{1}{2}}k}[\f{\pm8i\mu+(3-4\mu^2)\cn\,z\,\dn\,z}{\sn^2z}+2k\ln(\dn\,z-k\,\cn\,z)]\nonumber\\
&+\f{1}{2^6\alpha k^2}[\f{12-32\mu^2\pm i(38\mu-8\mu^3)\cn\,z\,\dn\,z}{\sn^4z}+\f{(1+k^2)(3-4\mu^2)(\pm i\mu\,\cn\,z\,\dn\,z-2)}{\sn^2z}]\nonumber\\
&+\cdots
\Big). \label{lame1stsmallwf}
\end{align}
Only in the case when all quantities, including the elliptic modulus $k$, take real values we have $\psi_\pm^*(z)=\psi_\mp(z)$. In particular, up to the first two leading order the wave functions can be written as
\begin{equation}
\psi_\pm(z)\sim\Big(\f{\dn\,z-k\,\cn\,z}{\dn\,z+k\,\cn\,z}\Big)^{\pm\f{\sqrt{\alpha}}{2}}\f{(\dn\,z\mp\cn\,z)^{-\f{i\mu}{2}-\f{1}{4}}}{(\dn\,z\pm\cn\,z)^{-\f{i\mu}{2}+\f{1}{4}}}.\label{lamesmallwavefucanother1}
\end{equation}
This asymptotic solution can be compared to the earlier results about the asymptotic Lam\'{e} function obtained by S. Malurkar in the 1930s, and results by H. M\"{u}ller in the 1960s \cite{Muller1966, MullerKirstenQUANT}.

In the limit $\alpha\to\infty,k\to0, \mu\to\nu$ with $\alpha^{\f{1}{2}}k\to 2ih^{\f{1}{2}}$ finite, we recover the unnormalized wave functions which differ some constant terms in the exponent from the asymptotic Mathieu wave functions (\ref{mathieu1stsmallwf}).

\vspace{3mm}
{\bf The second small energy expansion}

Around $z_*=\mathbf{K}$ we have $\Lambda=-\alpha k^2+\m{\em{small correction}}$, we set $\Lambda=-\alpha k^2+\widetilde{\Lambda}$ where $\widetilde{\Lambda}$ is a small quantity compared with the potential strength $\alpha k^2$. Here $\frac{1}{\sqrt{\alpha}}$ serves as the  expansion parameter. The relation $v_z+v^2=u-\alpha k^2+\widetilde{\Lambda}$ has an asymptotic solution in the form
\begin{equation}
v(z)=i\sum_{\ell=-1}^{\infty}\f{v_\ell(z)}{(\sqrt{\alpha})^\ell},
\end{equation}
with
\begin{subequations}
\begin{align}
v_{-1}&=k\,\cn\,z, \\
v_0&=\f{i}{2}\p_z\ln\cn\,z,\\
v_1&=\f{1}{8}k\,\cn\,z-\f{4\widetilde{\Lambda}+1-2k^2}{8k\,\cn\,z}+\f{3k^{\,\prime\,2}}{8k\,\cn^3z},\\
v_2&=-\f{i}{16k^2}\p_z\Big(\f{4\widetilde{\Lambda}+1-2k^2}{\cn^2z}-\f{3k^{\,\prime\,2}}{\cn^4z}\Big),\quad\m{etc},
\end{align}
\end{subequations}
where $k^{\,\prime}=\sqrt{1-k^2}$ is the complementary modulus.
After performing integration about the Jacobian elliptic functions we get the wave functions, now including both $\pm\sqrt{\alpha}$ sectors,
\begin{align}
\psi_\pm(z)=&\exp\Big(\pm\alpha^{\f{1}{2}}\ln(\dn\,z+ik\,\sn\,z)-\f{1}{2}\ln\cn\,z\nonumber\\
&\pm\f{i}{2^4\alpha^{\f{1}{2}}k}[\f{3\sn\,z\,\dn\,z}{\cn^2z}-2ik\ln(\dn\,z+ik\,\sn\,z)-\f{8\widetilde{\Lambda}-1+2k^2}{k^{\,\prime}}\ln\f{\dn\,z+k^{\,\prime}\sn\,z}{\cn\,z}]\nonumber\\
&-\f{3k^{\,\prime\,2}-(4\widetilde{\Lambda}+1-2k^2)\cn^2z}{2^4\alpha k^2\cn^4z}+\cdots
\Big).
\end{align}

The corresponding dispersion relation has been missed for a long time in the literature, motivated by some ideas from quantum gauge theory \cite{NS0908} recently we have derived it by the WKB analysis and a duality argument \cite{wh1108}, then we rederive it using the method adopted in this paper \cite{wh1412}. It is
\begin{align}
\widetilde{\Lambda}=&i2\alpha^{\f{1}{2}}k\mu+\f{1}{2^3}(1-2k^2)(\f{4\mu^2}{k^{\,\prime\,2}}+1)\nonumber\\
&+\f{i}{2^5\alpha^{\f{1}{2}}k}[\f{(1-2k^2)^2}{k^{\,\prime}}(\f{4\mu^3}{k^{\,\prime\,3}}+\f{3\mu}{k^{\,\prime}})+4k^2k^{\,\prime}(\f{4\mu^3}{k^{\,\prime\,3}}+\f{5\mu}{k^{\,\prime}})]\nonumber\\
&-\f{1-2k^2}{2^{10}\alpha k^2k^{\,\prime\,2}}(\f{80\mu^4}{k^{\,\prime\,4}}+\f{136\mu^2}{k^{\,\prime\,2}}+9)+\cdots.
\label{ellipsoidalJaceigenvalue3}
\end{align}
It gives us new asymptotic wave functions,
\begin{align}
\psi_\pm(z)=&\exp\Big(\pm\alpha^{\f{1}{2}}\ln(\dn\,z+ik\,\sn\,z)
-\f{1}{2}(\ln\cn\,z\mp\f{2\mu}{k^{\,\prime}}\ln\f{\dn\,z+k^{\,\prime}\sn\,z}{\cn\,z})\nonumber\\
&\pm\f{1}{2^4\alpha^{\f{1}{2}}k}[\f{\pm8ik^{\,\prime\,2}\mu+i(3k^{\,\prime\,2}+4\mu^2)\sn\,z\,\dn\,z}{k^{\,\prime\,2}\cn^2z}+2k\ln(\dn\,z+ik\,\sn\,z)]\nonumber\\
&-\f{1}{2^6\alpha k^2}[\f{k^{\,\prime\,2}(12k^{\,\prime\,2}+32\mu^2)\pm(38k^{\,\prime\,2}\mu+8\mu^3)\sn\,z\,\dn\,z}{k^{\,\prime\,2}\cn^4z}\nonumber\\
&\qquad\qquad+\f{(1-2k^2)(3k^{\,\prime\,2}+4\mu^2)(\pm\mu\,\sn\,z\,\dn\,z-2k^{\,\prime\,2})}{k^{\,\prime\,4}\cn^2z}]+\cdots
\Big).\label{lame2ndsmallwf}
\end{align}
They satisfy the property $\psi_\pm(-z)=\psi_\mp(z)$.
In companion with the expression (\ref{lamesmallwavefucanother1}) we could write the first two leading order results of the wave functions as
\begin{equation}
\psi_\pm(z)\sim\Big(\f{\dn\,z+ik\,\sn\,z}{\dn\,z-ik\,\sn\,z}\Big)^{\pm\f{\sqrt{\alpha}}{2}}\f{(\dn\,z\mp k^{\,\prime}\sn\,z)^{-\f{\mu}{2k^{\,\prime}}-\f{1}{4}}}{(\dn\,z\pm k^{\,\prime}\sn\,z)^{-\f{\mu}{2k^{\,\prime}}+\f{1}{4}}}.
\end{equation}
Taking the limit to the Mathieu wave functions (\ref{mathieu2ndsmallwf}), we would again encounter the difference of some constant terms which can be absorbed into the normalization constant.

Up to now, everything about the small energy expansions for the Lam\'{e} equation is consistent with the known results, although the monodromy relations, formulae (26) and (32) in Ref.~\cite{wh1412}, used to derive the corresponding dispersion relation remain  a physics induced conjecture.

\section{A connection to N=2 gauge theory}\label{connectoN=2}

Now we come back to the original motivation which inspired our study the spectral problem of periodic Schr\"{o}dinger operators, especially for the elliptic potentials. As we have shown in Refs.~\cite{wh1108,wh1412}, the asymptotic {\em eigenvalues} of the Mathieu and the Lam\'{e} equations are related to the solution of some deformed N=2 supersymmetric Yang-Mills gauge theories in the Nekrasov-Shatashvili limit (NS) \cite{NS0908}. The three asymptotic spectral solutions are precisely in accordance with three different dual descriptions of the low energy effective physics of gauge theory, i.e. the Seiberg-Witten duality \cite{SW9407,SW9408}, in particular the large energy solution is related to the Nekrasov instanton partition function \cite{instcount}.

The large energy asymptotic {\em wave functions} are related to the instanton partition function of gauge theory with surface operator inserted. The partition function with surface operator extends Nekrasov's localization formula, it is introduced and developed in Refs.~\cite{Braverman0401, BravermanEtingof0409}. The computation can be carried out by the characters developed in Ref.~\cite{FFNR0812}. The paper by Alday and Tachikawa gives a detailed study about the relations between the SU(2) gauge theory with surface operator, the SL(2) conformal block and the two-body quantum Calogero-Moser model \cite{at1005}. In the following, we briefly explain the relation between the gauge theory partition function with surface operator and the asymptotic wave functions (\ref{largeenergywavefunctionmathieu}) and (\ref{largeenergywavefunctionlame}).

Let us start from the SU(2) N=2$^*$ gauge theory with surface operator, whose partition function takes the following form,
\begin{equation} Z(a,m,\epsilon_{1,2},x_{1,2})=\sum_{k_1,k_2\geqslant0}^{\infty}Z_{k_1,k_2}(a,m,\epsilon_{1,2})x_1^{k_1}x_2^{k_2},
\end{equation}
where $a$ is the scalar v.e.v, $m$ is the mass of adjoint matter, $\epsilon_1,\epsilon_2$ are the $\Omega$-deformation and $x_1,x_2$ are the counting parameters. Written in the exponential form, its pole structure in the limit $\epsilon_1\to0,\epsilon_2\to0$ is
\begin{equation}
Z=\exp(\f{F}{\epsilon_1\epsilon_2}+\f{G}{\epsilon_1}).\label{polesofZ}
\end{equation}
The functions $F$ and $G$ are $x_1,x_2$-asymptotic series, their coefficients are functions of $a,m, \epsilon_1,\epsilon_2$ and in particular are regular with respect to $\epsilon_1,\epsilon_2$. There remains an ambiguity of distributing terms, because a term like $\f{\mathbf{f}(a,m,x_{1,2})}{\epsilon_1}$ can be put in either $\f{F}{\epsilon_1\epsilon_2}$ or $\f{G}{\epsilon_1}$. To fix the ambiguity we require that the function $F$ contains only terms with expansion parameters of the form $(x_1x_2)^n$, while $G$ does not contain such terms. The presence of the surface operator beaks the symmetry between $\epsilon_1$ and $\epsilon_2$ in $F$ and $G$, therefore $F$ is not exactly the deformed prepotential obtained from the Nekrasov partition function, nevertheless, in the NS limit $\epsilon_2\to0$ they give the same eigenvalue as in (\ref{lambdafromF}).

In order to relate gauge theory and the quantum mechanics spectral problem, some manipulations on the function $Z$ are needed. The spectral solution of the Lam\'{e} operator is related to the large-$a$-expansion of instanton partition function (\ref{polesofZ}), in accordance with the large-$\nu$-expansions of the eigenvalue (\ref{largeenergyeigenvaluelame}) and the eigenfunctions  (\ref{largeenergywavefunctionlame}). Both $\f{F}{\epsilon_1\epsilon_2}$ and $\f{G}{\epsilon_1}$ contain $a$-independent terms when expanded as large-$a$-series, which deserve special attention. These terms are polynomials of $x_1,x_2$, and can be represented by the Dedekind eta function and the elliptic theta function,
\begin{align}
\f{F}{\epsilon_1\epsilon_2}&=\f{2(m-\epsilon_1)(m-\epsilon_2)}{\epsilon_1\epsilon_2}\ln\Big(\f{(x_1x_2)^{\f{1}{24}}}{\eta(x_1x_2)}\Big)+\mathcal {O}(a^{-2}),\\
\f{G}{\epsilon_1}&=\f{m-\epsilon_1}{\epsilon_1}\ln\Big[\vartheta_4\Big(\f{i}{4}\ln\f{x_1}{x_2},x_1x_2\Big)\f{(x_1x_2)^{\f{1}{24}}}{\eta(x_1x_2)}\Big]+\mathcal {O}(a^{-1}).
\end{align}

To see the connections of functions $F, G$ and the eigenvalue, eigenfunction, we first need to identify the parameters by
\begin{equation}
\f{\pi a}{\epsilon_1}=\omega_1\nu,\quad \f{m}{\epsilon_1}=n,\quad x_1=q^{\f{1}{2}}e^{-\f{i\pi}{\omega_1}x},\quad x_2=q^{\f{1}{2}}e^{\f{i\pi}{\omega_1}x}.\label{parametertransfor}
\end{equation}
The elliptic nome $q$ is the instanton parameter of gauge theory, therefore, the function $F$ is a $x$-independent $q$-series which gives the eigenvalue, the function $G$ is a $q$-series depending on the coordinate $x$ which gives the wave function. The eigenvalue $\lambda$ in (\ref{largeenergyeigenvaluelame}) is related to the function $F$ in the limit $\epsilon_2\to0$ by
\begin{align}
\lambda&=-\nu^2+\f{\pi^2}{\omega_1^2}\Big[q\f{\p}{\p q}\f{F|_{\epsilon_2\to0}}{\epsilon_1^2}-\f{\alpha}{12}(1-E_2)\Big]\nonumber\\
&=-\nu^2+\f{\pi^2}{\omega_1^2}\Big[q\f{\p}{\p q}\Big(2\alpha\ln\Big(\f{q^{\f{1}{24}}}{\eta(q)}\Big)-\f{\pi^2}{\omega_1^2}\f{\alpha^2(q+3q^2+4q^3+\cdots)}{2\nu^2}+\cdots\Big)-\f{\alpha}{12}(1-E_2)\Big]\nonumber\\
&=-\nu^2+\f{\alpha^2(12\zeta_1^2-g_2)}{48\nu^2}+\mathcal{O}(\nu^{-4}).\label{lambdafromF}
\end{align}
In gauge theory the term $-\nu^2$ is perturbative, hence not included in the instanton partition function. This relation is examined in detail in Ref.~\cite{wh1108} (see formula (34) in that paper), there is a difference of $\alpha\zeta_1=\alpha\f{\pi^2}{12\omega_1^2}E_2$ on the right hand side because here we use the shifted potential $\widetilde{\wp}(x)$.

On the other hand, the wave functions (\ref{largeenergywavefunctionlame}) is related to the function $G$ in the limit $\epsilon_2\to0$ by
\begin{equation}
e^{-i\nu(x+\omega_2)}\psi_{\pm}(\pm x\pm\omega_2,\nu,\alpha,q)\Big[\vartheta_4(\f{\pi x}{2\omega_1},q)\f{q^{\f{1}{24}}}{\eta(q)}\Big]^{n-1}
=\exp[\f{G(a,m,\epsilon_{1,2},x_{1,2})}{\epsilon_1}]|_{\epsilon_2\to0}.\label{wavefunctionandG}
\end{equation}
In the expression we emphasize the parameters used on both sides, and use the property of the large energy  wave functions $\psi_-(-x-\omega_2)=\psi_+(x+\omega_2)$.
For example, up to the order $a^{-2}$ we have
\begin{align}
\f{G(\epsilon_2=0)}{\epsilon_1}=&-\f{m-\epsilon_1}{\epsilon_1}(x_1+x_2+\f{1}{2}x_1^2+\f{1}{2}x_2^2+\f{1}{3}x_1^3+x_1^2x_2+x_1x_2^2+\f{1}{3}x_2^3+\cdots)\nonumber\\
&-\f{m(m-\epsilon_1)}{2a\epsilon_1}(x_1-x_2+x_1^2-x_2^2+x_1^3+x_1^2x_2-x_1x_2^2-x_2^3+\cdots)\nonumber\\
&-\f{m(m-\epsilon_1)}{4a^2}(x_1+x_2+2x_1^2+2x_2^2+3x_1^3+x_1^2x_2+x_1x_2^2+3x_2^3+\cdots)\nonumber\\
&-\mathcal {O}(a^{-3}).
\label {Gexpanded}
\end{align}
Using the relation of parameters given in (\ref{parametertransfor}),
the first three pieces are summed into  three elliptic functions,
\begin{equation}
(n-1)\ln\Big[\vartheta_4(\f{\pi
x}{2\omega_1},q)\f{q^{\f{1}{24}}}{\eta(q)}\Big],\qquad
\f{i\alpha\p_x\ln\vartheta_4(\f{\pi
x}{2\omega_1},q)}{2\nu},\qquad-\f{\alpha\p_x^2\ln\vartheta_4(\f{\pi
x}{2\omega_1},q)}{4\nu^2},
\end{equation} which precisely match the left hand
side of (\ref{wavefunctionandG}). The coefficients of the term
$\nu^{-l}$, with $l\geqslant1$, are linear
polynomials of $\p_x^k\ln\vartheta_4(\f{\pi x}{2\omega_1},q)$ with
$k\geqslant1$. As the argument of wave functions
in (\ref{wavefunctionandG}) is $x+\omega_2$, we have
$\widetilde{\zeta}(x+\omega_2)=\p_x\ln\vartheta_4(\f{\pi x}{2\omega_1},q)$, therefore, the coefficients of the term
$\nu^{-l}$ can be rewritten as linear polynomials of $\p_x^k\widetilde{\zeta}(x+\omega_2)$ with $k\geqslant0$, in accordance with the discussion in \ref{lamelargeenergy}.
In another form, these coefficients can be rewritten as linear polynomials of the Jacobi zeta function and its derivatives, because $\p_x\ln\vartheta_4(\f{\pi x}{2\omega_1},q)=\f{\mathbf{K}}{\omega_1}\mbox{zn}(\f{\mathbf{K}x}{\omega_1}|k^2)$.

The eigenvalue (\ref{largeenergyeigenvaluelame}) and the eigenfunction (\ref{largeenergywavefunctionlame}) provide an elliptic modular representation for the gauge theory partition function when $\epsilon_2=0$.
In fact, we observe evidence that even for the case when both deformation parameters are turned on, $\epsilon_1\ne0,\epsilon_2\ne0$, the instanton partition function with surface operator can be expressed in terms of theta functions. This property indicates the instanton partition function secretly records relations to the  elliptic curve. Indeed, this connection can be seen from the point of view of either integrable system \cite{NS0908} or conformal field theory \cite{at1005, agt}.

In the decoupling limit, the N=2$^*$ gauge theory becomes the pure gauge theory. The corresponding partition function with surface operator can be found in Ref.~\cite{AFKMY1008}, it is related to the asymptotic Mathieu wave functions (\ref{largeenergywavefunctionmathieu}).

\appendix

\section{A matrix that counts divisors of integers}

When we take the $a\to\infty$ limit of the instanton partition with surface operator,
only $a$-independent terms in the functions $F$ and $G$ remain. These terms are represented by two elliptic modular functions,
\begin{align}
\f{\eta(q)}{q^{\f{1}{24}}}&=\prod_{n=1}^{\infty}(1-q^n)=\sum_{n=-\infty}^{\infty}(-1)^nq^{\f{n(3n-1)}{2}},\\
\vartheta_4(\chi,q)&=\prod_{n=1}^{\infty}(1-2q^{n-\f{1}{2}}\cos2\chi+q^{2n-1})(1-q^n)\nonumber\\
 &=1+2\sum_{n=1}^{\infty}(-1)^{n}q^{\f{n^2}{2}}\cos2n\chi,
\end{align}
with $\chi=\f{\pi x}{2\omega_1}$. Had we expanded them as $q$-series as usual, there might not be interesting things deserve to say. Nevertheless, if we rewrite them in terms of $x_1,x_2$ as given in (\ref{parametertransfor}), and then expand minus of the logarithm of them as series of $x_1,x_2$, we get
\begin{align}
-\ln\Big(\f{\eta(x_1x_2)}{(x_1x_2)^{\f{1}{24}}}\Big)=&x_1x_2+\f{3}{2}x_1^2x_2^2+\f{4}{3}x_1^3x_2^3+\f{7}{4}x_1^4x_2^4+\f{6}{5}x_1^5x_2^5+2x_1^6x_2^6+\cdots,\label{lneta}\\
-\ln\vartheta_4\Big(\f{i}{4}\ln\f{x_1}{x_2},x_1x_2\Big)=&x_1x_2+\f{3}{2}x_1^2x_2^2+\f{4}{3}x_1^3x_2^3+\f{7}{4}x_1^4x_2^4+\f{6}{5}x_1^5x_2^5+2x_1^6x_2^6+\cdots\nonumber\\
&+x_1+x_2+\f{1}{2}x_1^2+\f{1}{2}x_2^2+\f{1}{3}x_1^3+x_1^2x_2+x_1x_2^2+\f{1}{3}x_2^3\nonumber\\
&+\f{1}{4}x_1^4+\f{1}{4}x_2^4+\f{1}{5}x_1^5+x_1^3x_2^2+x_1^2x_2^3+\f{1}{5}x_2^5\nonumber\\
&+\f{1}{6}x_1^6+\f{1}{2}x_1^4x_2^2+\f{1}{2}x_1^2x_2^4+\f{1}{6}x_2^6+\cdots=\sum_{i=0,j=0}^{\infty}\Theta_4[i,j]x_1^ix_2^j.\label{lntheta4}
\end{align}
The coefficient matrix $\Theta_4[i,j]$ is a symmetric infinite matrix with all elements positive, as a digest here we present the first 22 dimensions. Notice that the numbers for rows and columns of the matrix begin from 0.

\begin{small}
\begin{equation*}
\left(
\begin{array}{cccccccccccccccccccccc}
 0 & 1 & \frac{1}{2} & \frac{1}{3} & \frac{1}{4} & \frac{1}{5} & \frac{1}{6} & \frac{1}{7} & \frac{1}{8} & \frac{1}{9} & \frac{1}{10} & \frac{1}{11} & \frac{1}{12} & \frac{1}{13} & \frac{1}{14} & \frac{1}{15} & \frac{1}{16} & \frac{1}{17} & \frac{1}{18} & \frac{1}{19} & \frac{1}{20} & \frac{1}{21} \\
 1 & 1 & 1 & 0 & 0 & 0 & 0 & 0 & 0 & 0 & 0 & 0 & 0 & 0 & 0 & 0 & 0 & 0 & 0 & 0 & 0 & 0 \\
 \frac{1}{2} & 1 & \frac{3}{2} & 1 & \frac{1}{2} & 0 & 0 & 0 & 0 & 0 & 0 & 0 & 0 & 0 & 0 & 0 & 0 & 0 & 0 & 0 & 0 & 0 \\
 \frac{1}{3} & 0 & 1 & \frac{4}{3} & 1 & 0 & \frac{1}{3} & 0 & 0 & 0 & 0 & 0 & 0 & 0 & 0 & 0 & 0 & 0 & 0 & 0 & 0 & 0 \\
 \frac{1}{4} & 0 & \frac{1}{2} & 1 & \frac{7}{4} & 1 & \frac{1}{2} & 0 & \frac{1}{4} & 0 & 0 & 0 & 0 & 0 & 0 & 0 & 0 & 0 & 0 & 0 & 0 & 0 \\
 \frac{1}{5} & 0 & 0 & 0 & 1 & \frac{6}{5} & 1 & 0 & 0 & 0 & \frac{1}{5} & 0 & 0 & 0 & 0 & 0 & 0 & 0 & 0 & 0 & 0 & 0 \\
 \frac{1}{6} & 0 & 0 & \frac{1}{3} & \frac{1}{2} & 1 & 2 & 1 & \frac{1}{2} & \frac{1}{3} & 0 & 0 & \frac{1}{6} & 0 & 0 & 0 & 0 & 0 & 0 & 0 & 0 & 0 \\
 \frac{1}{7} & 0 & 0 & 0 & 0 & 0 & 1 & \frac{8}{7} & 1 & 0 & 0 & 0 & 0 & 0 & \frac{1}{7} & 0 & 0 & 0 & 0 & 0 & 0 & 0 \\
 \frac{1}{8} & 0 & 0 & 0 & \frac{1}{4} & 0 & \frac{1}{2} & 1 & \frac{15}{8} & 1 & \frac{1}{2} & 0 & \frac{1}{4} & 0 & 0 & 0 & \frac{1}{8} & 0 & 0 & 0 & 0 & 0 \\
 \frac{1}{9} & 0 & 0 & 0 & 0 & 0 & \frac{1}{3} & 0 & 1 & \frac{13}{9} & 1 & 0 & \frac{1}{3} & 0 & 0 & 0 & 0 & 0 & \frac{1}{9} & 0 & 0 & 0 \\
 \frac{1}{10} & 0 & 0 & 0 & 0 & \frac{1}{5} & 0 & 0 & \frac{1}{2} & 1 & \frac{9}{5} & 1 & \frac{1}{2} & 0 & 0 & \frac{1}{5} & 0 & 0 & 0 & 0 & \frac{1}{10} & 0 \\
 \frac{1}{11} & 0 & 0 & 0 & 0 & 0 & 0 & 0 & 0 & 0 & 1 & \frac{12}{11} & 1 & 0 & 0 & 0 & 0 & 0 & 0 & 0 & 0 & 0 \\
 \frac{1}{12} & 0 & 0 & 0 & 0 & 0 & \frac{1}{6} & 0 & \frac{1}{4} & \frac{1}{3} & \frac{1}{2} & 1 & \frac{7}{3} & 1 & \frac{1}{2} & \frac{1}{3} & \frac{1}{4} & 0 & \frac{1}{6} & 0 & 0 & 0 \\
 \frac{1}{13} & 0 & 0 & 0 & 0 & 0 & 0 & 0 & 0 & 0 & 0 & 0 & 1 & \frac{14}{13} & 1 & 0 & 0 & 0 & 0 & 0 & 0 & 0 \\
 \frac{1}{14} & 0 & 0 & 0 & 0 & 0 & 0 & \frac{1}{7} & 0 & 0 & 0 & 0 & \frac{1}{2} & 1 & \frac{12}{7} & 1 & \frac{1}{2} & 0 & 0 & 0 & 0 & \frac{1}{7} \\
 \frac{1}{15} & 0 & 0 & 0 & 0 & 0 & 0 & 0 & 0 & 0 & \frac{1}{5} & 0 & \frac{1}{3} & 0 & 1 & \frac{8}{5} & 1 & 0 & \frac{1}{3} & 0 & \frac{1}{5} & 0 \\
 \frac{1}{16} & 0 & 0 & 0 & 0 & 0 & 0 & 0 & \frac{1}{8} & 0 & 0 & 0 & \frac{1}{4} & 0 & \frac{1}{2} & 1 & \frac{31}{16} & 1 & \frac{1}{2} & 0 & \frac{1}{4} & 0 \\
 \frac{1}{17} & 0 & 0 & 0 & 0 & 0 & 0 & 0 & 0 & 0 & 0 & 0 & 0 & 0 & 0 & 0 & 1 & \frac{18}{17} & 1 & 0 & 0 & 0 \\
 \frac{1}{18} & 0 & 0 & 0 & 0 & 0 & 0 & 0 & 0 & \frac{1}{9} & 0 & 0 & \frac{1}{6} & 0 & 0 & \frac{1}{3} & \frac{1}{2} & 1 & \frac{13}{6} & 1 & \frac{1}{2} & \frac{1}{3} \\
 \frac{1}{19} & 0 & 0 & 0 & 0 & 0 & 0 & 0 & 0 & 0 & 0 & 0 & 0 & 0 & 0 & 0 & 0 & 0 & 1 & \frac{20}{19} & 1 & 0 \\
 \frac{1}{20} & 0 & 0 & 0 & 0 & 0 & 0 & 0 & 0 & 0 & \frac{1}{10} & 0 & 0 & 0 & 0 & \frac{1}{5} & \frac{1}{4} & 0 & \frac{1}{2} & 1 & \frac{21}{10} & 1 \\
 \frac{1}{21} & 0 & 0 & 0 & 0 & 0 & 0 & 0 & 0 & 0 & 0 & 0 & 0 & 0 & \frac{1}{7} & 0 & 0 & 0 & \frac{1}{3} & 0 & 1 & \frac{32}{21}
\end{array}
\right)
\end{equation*}
\end{small}

A gaze reveals that the matrix $\Theta_4[i,j]$ contains information about an elementary fact of number theory. Given a positive integer $n\geqslant1$, there are $d(n)$ positive divisors denoted as the set  $\{d_1,d_2,d_3,\cdots d_{d(n)}\}$, in a decreasing order $d_1>d_2>\cdots>d_{d(n)}$. Among the divisors, we have $d_1=n, d_{d(n)}=1$. The divisor function is defined by $\sigma_k(n)=\sum\limits_{i}d_i^k$.
The nonzero diagonal elements of the matrix, which are the coefficients of the $q$-series for   $\ln\f{q^{\f{1}{24}}}{\eta(q)}$, are $\sigma_{-1}(n)$ for all positive integers, $\Theta_4[n,n]=\sigma_{-1}(n)$. Then the expansion of the logarithm of eta function is  $\ln\eta(q)=\f{1}{24}\ln q-\sum\limits_{n=1}^\infty\sigma_{-1}(n)q^n$. A further derivative gives the expression for the second Eisenstein series, $E_2(q)=24q\p_q\ln\eta(q)=1-24\sum\limits_{n=1}^\infty\sigma_1(n)q^n$, where the relation $n\sigma_{-1}(n)=\sigma_1(n)$ is used. The nonzero off-diagonal elements of the the matrix are precisely the inverse of the divisors of integers. If we focus on the lower triangular part, in the $n$-th row  the nonzero numbers are $\{d_1^{-1},d_2^{-1},d_3^{-1},\cdots d_{d(n)}^{-1}\}$ and their sum:
\begin{equation}
\sum_{i=1}^{d(n)}\f{1}{d_i}=\Theta_4[n,n].
\end{equation}
For example, in the $18$-th row we have the divisor function $\sigma_{-1}(18)$,
\begin{equation}
\f{1}{18}+\f{1}{9}+\f{1}{6}+\f{1}{3}+\f{1}{2}+\f{1}{1}=\f{13}{6}.
\end{equation}
This fact indicates a relation of the eta function and the theta function expanded as in (\ref{lneta}),(\ref{lntheta4}):
while the logarithm of eta function knows the infinite sequence of numbers $1,\f{3}{2},\f{4}{3},\f{7}{4},\cdots$, the logarithm of theta function diagnoses where they come from.
It is the same in spirit for the situation in gauge theory: while the instanton partition function {\em without} surface operator knows the eigenvalue expansion, it is the instanton partition function {\em with} surface operator tells the whole story.

If a divisor $d_i$ is a composite number, we can always move up to the $d_i$-th row to find its divisors other than 1 and $d_i$. Eventually, we could arrive at the unique prime factorization for the integer $n=\prod\limits_{i=1}^sp_i^{\nu_i}$. A well known fact of number theory asserts that the number of nonzero elements from $\Theta_4[n,0]$ to $\Theta_4[n,n-1]$ in the $n$-th row is given by $d(n)=\prod\limits_{i=1}^s(\nu_i+1)$. The $n$-th column tells the same story.

In fact, the story goes a bit further. The large $\nu$ expansion of wave function (instanton partition function) involves higher order $\chi$-derivatives of $\ln\vartheta_4(\chi,q)$.
All elliptic functions in (\ref{wavefunctionandG}), (\ref{Gexpanded}) are made out of $\p_\chi^k\ln\vartheta_4(\chi,q)$, with $k\geqslant0$, they count the
$(k-1)$-th power of divisors. More precisely, the coefficient matrix of $-(\f{i}{2})^k\p_\chi^k\ln\vartheta_4(\chi,q)$ expanded as $x_1,x_2$-series counts $d_i^{k-1}$ for all positive integers. The discussion above is just about the first case $k=0$.

This might be a folklore of number theory, nevertheless, it is strange instanton knows it.

\section*{Acknowledgments}

I thank Institute of Modern Physics at Northwest University, Xi'an, during the summer school ``Integrable models and their applications 2016" where this work is finalized. Part of the work has been done when I was supported by the FAPESP No. 2011/21812-8, through IFT-UNESP.


\begin{thebibliography}{10}

\bibitem{wh1412}
W. He,
\newblock A new treatment for some Schr\"{o}dinger operators I: the eigenvalue.
\newblock {Commun. Theor. Phys.} {\bf 69} (2018) 115--126.
\newblock \href{http://arxiv.org/abs/1412.6776}{{\tt arXiv:1412.6776[math-ph]}}.

\bibitem{NS0908}
N.~Nekrasov and S.~Shatashvili,
\newblock Quantization of integrable systems and four dimensional gauge theories,
\newblock In {\em 16th International Congress on Mathematical Physics}, World Scientific, Singapore (2010), 265-289.
\newblock \href{http://arxiv.org/abs/0908.4052}{{\tt arXiv:0908.4052[hep-th]}}.

\bibitem{MGK}
R. M. Miura, C. S. Gardner, and M. D. Kruskal,
\newblock Korteweg-de Vries Equation and Generalizations. II. Existence of Conservation Laws and Constants of Motion,
\newblock  J. Math. Phys. {\bf 9} (1968) 1204.

\bibitem{classintegrable}
O.~Babelon, D.~Bernard, and M.~Talon,
\newblock {\em Introduction to Classical Integrable Systems},
\newblock Cambridge University Press, Cambridge (2003).

\bibitem{mclachlan}
N.~W. McLachlan,
\newblock {\em Theory and application of Mathieu functions},
\newblock Oxford University Press, Oxford (1947).

\bibitem{Arscott1964}
F.~M. Arscott,
\newblock {\em Periodic differential equations},
\newblock Pergamon Press, Oxford (1964).

\bibitem{Wang-Guo}
Z.-X. Wang and D.-R. Guo,
\newblock {\em Special Functions},
\newblock World Scientific, Singapore (1989).

\bibitem{NIST}
{\it NIST Digital Library of Mathematical Functions.}
F. W. J. Olver, A. B. Olde Daalhuis, D. W. Lozier, B. I. Schneider, R. F. Boisvert, C. W. Clark, B. R. Miller, and B. V. Saunders, eds.
\newblock \href{http://dlmf.nist.gov/}{{\tt http://dlmf.nist.gov/}}.

\bibitem{Langmann2004b}
E.~Langmann,
\newblock An explicit solution of the (quantum) elliptic Calogero-Sutherland model,
\newblock In {\em Symmetry and Perturbation Theory (Cala Gonone)}, World Scientific, Singapore (2005), 159--174.
\newblock \href{http://arxiv.org/abs/math-ph/0407050}{{\tt arXiv:math-ph/0407050}}.


\bibitem{wh1108}
W.~He,
\newblock Combinatorial approach to Mathieu and Lam\'{e} equations,
\newblock J. Math. Phys. {\bf 56} (2015) 072302 .
\newblock \href{http://arxiv.org/abs/1108.0300}{{\tt arXiv:1108.0300[math-ph]}}.

\bibitem{wh1401}
W.~He,
\newblock Quasimodular instanton partition function and the elliptic solution of Korteweg-de Vries equations,
\newblock Ann. Phys. {\bf 353} (2015) 150--162.
\newblock \href{http://arxiv.org/abs/1401.4135}{{\tt arXiv:1401.4135[hep-th]}}.

\bibitem{Stone-Reeve1978}
M.~Stoner and J.~Reeve,
\newblock Late terms in the asymptotic expansional for the energy levels of a periodic potential,
\newblock Phys. Rev. D {\bf 18} (1978) 4746.

\bibitem{MullerKirstenQUANT}
H.~J.~W. M\"{u}ller-Kirsten,
\newblock {\em Introduction to quantum mechanics: Schr\"{o}dinger equation and path integral},
\newblock World Scientific, Singapore (2006).

\bibitem{Dingle-Muller1962}
R.~B. Dingle and H.~J.~W. M\"{u}ller,
\newblock Asymptotic Expansions of Mathieu Functions and their Characteristic Numbers,
\newblock J. reine und angew. Math.  {\bf 211} (1962) 11--32.

\bibitem{Muller1966}
H.~J.~W. M\"{u}ller,
\newblock Asymptotic Expansions of Ellipsoidal Wave Functions and their Characteristic Numbers,
\newblock  Math. Nachr. {\bf 31} (1966) 89--101.

\bibitem{SW9407}
N.~Seiberg and E.~Witten,
\newblock Electric-Magnetic Duality, Monopole Condensation, And Confinement In N=2 Supersymmetric Yang-Mills Theory,
\newblock Nucl. Phys. B {\bf 426} (1994) 19--52.
\newblock \href{http://arxiv.org/abs/hep-th/9407087}{{\tt arXiv:hep-th/9407087}}.

\bibitem{SW9408}
N.~Seiberg and E.~Witten,
\newblock Monopoles, Duality and Chiral Symmetry Breaking in N=2 Supersymmetric QCD,
\newblock Nucl. Phys. B {\bf 431} (1994) 484--550.
\newblock \href{http://arxiv.org/abs/hep-th/9408099}{{\tt arXiv:hep-th/9408099}}.

\bibitem{instcount}
N.~Nekrasov,
\newblock Seiberg-Witten Prepotential From Instanton Counting,
\newblock Adv. Theor. Math. Phys. {\bf 7} (2004) 831--864.
\newblock \href{http://arxiv.org/abs/hep-th/0206161}{{\tt arXiv:hep-th/0206161}}.

\bibitem{Braverman0401}
A.~Braverman,
\newblock Instanton counting via affine Lie algebras I: Equivariant J-functions of (affine) flag manifolds and Whittaker vectors,
\newblock {\it Proceedings of the CRM Workshop on Algebraic Structures and Moduli Spaces (Montreal)}, American Mathematical Society, Providence (2004).
\newblock \href{http://arxiv.org/abs/math/0401409}{{\tt arXiv:math/0401409}}.

\bibitem{BravermanEtingof0409}
A.~Braverman and P.~Etingof,
\newblock Instanton counting via affine Lie algebras II: from Whittaker vectors to the Seiberg-Witten prepotential,
\newblock \href{http://arxiv.org/abs/math/0409441}{{\tt arXiv:math/0409441}}.

\bibitem{FFNR0812}
B.~Feigin, M.~Finkelberg, A.~Negut, and L.~Rybnikov,
\newblock Yangians and cohomology rings of Laumon spaces,
\newblock Selecta Mathematica {\bf 17} (2011) 513.
\newblock \href{http://arxiv.org/abs/0812.4656}{{\tt arXiv:0812.4656[math.AG]}}.

\bibitem{at1005}
L.~F. Alday and Y.~Tachikawa,
\newblock Affine {SL(2)} conformal blocks from 4d gauge theories,
\newblock Lett. Math. Phys.  {\bf 94} (2010) 87--114.
\newblock \href{http://arxiv.org/abs/1005.4469}{{\tt arXiv:1005.4469[hep-th]}}.

\bibitem{agt}
L.~F. Alday, D.~Gaiotto, and Y.~Tachikawa,
\newblock Liouville correlation functions from four dimensional gauge theories,
\newblock Lett. Math. Phys. {\bf 91} (2010) 167--197.
\newblock \href{http://arxiv.org/abs/0906.3219}{{\tt arXiv:0906.3219[hep-th]}}.

\bibitem{AFKMY1008}
H.~Awata, H.~Fuji, H.~Kanno, M.~Manabe, and Y.~Yamada,
\newblock Localization with a Surface Operator, Irregular Conformal Blocks and Open Topological String,
\newblock Adv. Theor. Math. Phys. {\bf 16} (2012) 725--804.
\newblock \href{http://arxiv.org/abs/1008.0574}{{\tt arXiv:1008.0574[hep-th]}}.

\end{thebibliography}
\end{document}